# Fiscal policy and inequality in a model with endogenous positional concerns

Kirill Borissov*and Nigar Hashimzade†

July 1, 2021

## Abstract

We investigate the dynamics of wealth inequality in an economy where households have positional preferences, with the strength of the positional concern determined endogenously by inequality of wealth distribution in the society. We demonstrate that in the long run such an economy converges to a unique egalitarian steady-state equilibrium, with all households holding equal positive wealth, when the initial inequality is sufficiently low. Otherwise, the steady state is characterised by polarisation of households into rich, who own all the wealth, and poor, whose wealth is zero. A fiscal policy with government consumption funded by taxes on labour income and wealth can move the economy from any initial state towards an egalitarian equilibrium with a higher aggregate wealth.

*Department of Economics, European University at St. Petersburg, 6/1A Gagarinskaya Str., St. Petersburg, 191187, Russia.
†Corresponding authors. Department of Economics and Finance, Brunel University London, Kingston Lane, Uxbridge, Middlesex, UB8 3PH, United Kingdom. E-mail: nigar.hashimzade@brunel.ac.uk

**JEL classification** D31, D91, E21, H31

# 1 Introduction

Social sciences have long recognised the importance of social comparisons for human behaviour. The idea that humans care about their position relative to other members of the society is, perhaps, as old as the human societies. In the political economy writings it was mentioned by Marx (1849) and Veblen (1899), and formalised by Duesenberry (1949). The contemporary economic literature has accumulated a large body of empirical and experimental evidence on the positional, or relative nature of individual preferences (Clark et al., 2008; Heffetz and Frank, 2011).

One common approach to modelling positional concern, referred to as 'keeping up with the Joneses' (Gali, 1994), is to assume that individual utility increases in own consumption, and, in addition, depends on consumption relative to some benchmark level, – often the average across the society or in the relevant reference group. An individual experiences utility gain, or relative satisfaction, from a positive gap between his or her consumption and the benchmark. Conversely, he or she experiences utility loss, or relative deprivation, from a negative consumption gap. The relative component is sometimes defined in terms of income gap or wealth gap.

There is also evidence in the empirical and experimental literature that individuals care about the distribution of income or wealth (Clark and D'Ambrosio, 2014). Distributional concerns are modelled as a negative or positive relationship between an individual's utility and the degree of inequality in the society or in the peer group. The attitudes to inequality may depend on which inequality is considered: there is an argument that people dislike inequality in unearned wealth, or endowments, and favour inequality in earned income, or rewards (Hopkins and Kornienko, 2010).

In this paper we model individual preferences in a dynamic interaction of the positional and distributional concerns by allowing an increase in wealth inequality to raise the importance of consumption relative to others. We use this framework to investigate how the distribution of wealth evolves over time and how a fiscal policy with government consumption funded by taxes can reduce wealth inequality.

Our analysis of the dynamic evolution of such an economy shows that the long-run outcome depends on the initial distribution of wealth. If the initial inequality is high, the economy converges to a polarised equilibrium where population divides into two classes, the poor with zero wealth and the rich who hold the entire wealth. Otherwise, if the initial inequality is sufficiently low, the economy converges to an egalitarian equilibrium, with aggregate wealth distributed equally among households. Moreover, with the same starting aggregate wealth, in the long run the aggregate output and wealth are higher in the egalitarian equilibrium than in any of the polarised equilibria.

In this setting we show that the adverse effect of the initial inequality on the long-run distribution of wealth can be overcome by implementing a simple fiscal policy. Specifically, we show that a set of taxes on labour income, capital income, and inherited wealth, with revenues used to fund public consumption, can lead the economy from an initial state with any degree of inequality to an egalitarian equilibrium in the long run. Furthermore, it is possible to construct a combination of taxes in such a way that in the long run there is no trade-off between the aggregate wealth and equality.

Our main assumption is to allow the strength of envy to depend on the wealth distribution. This is motivated by empirical observations in sociology suggesting that a higher place in social hierarchy is valued more in a more unequal society. There is evidence that in more unequal societies people experience higher stress and anxiety about their relative position in the society (Wilkinson and Pickett, 2006, 2010), care more about respect, admiration, and recognition by other people (Paskov et al., 2013), and devote more resources to acquisition of positional goods, such as designer brands and expensive jewellery (Walasek and Brown, 2015). Strong positive association between social comparisons and inequality was found in several empirical studies of life satisfaction. For example, Cheung and Lucas (2016) studied the role of income inequality as a moderator of the relative income effect on subjective wellbeing, using the Behavioral Risk Factor Surveillance System[1] data on 1.7 million individuals from 2,425 counties in the United States. They found that higher neighborhood income was associated with lower life satisfaction, and that social comparison of income was stronger in the counties with higher income inequality. Similarly, in an empirical analysis based on the Panel Study of Income Dynamics,[2] Brown et al. (2017) found that life satisfaction depends on the relative rank position within a social comparison group, and that the effects of relative rank are stronger when income inequality is high. Schneider (2019) argued that subjective social status is an important psychological mechanism that drives the link between inequality and life satisfaction. Using 2012/13 European Social Survey,[3] she found that income inequality, measured by the country-level Gini coefficient, increases the importance of subjective social status to life satisfaction. While these observations do not indicate the direction of causality, the positive association between inequality and envy is consistent with the model predictions, where high inequality coupled with strong positional concerns or low inequality coupled with weak positional concerns emerge endogenously in the long run.

---

[1] An annual telephone survey conducted since 1984 by the U.S. Center for Disease Control and Prevention and health departments of individual States. The survey tracks health information in the United States. See https://www.cdc.gov/brfss/.

[2] A longitudinal survey of a representative sample of U.S. households launched in 1968.The data cover numerous economic and socio-demographic topics. See https://psidonline.isr.umich.edu/.

## 2  Related Literature

Our work contributes to the literature on the aggregate and distributional outcomes of dynamic social preferences and the implications of social preferences for policy choices. While much of the theoretical literature on the role of positional and distributional concerns in economic outcomes and on the distributional consequences of positional preferences has focused on the models where the social preferences are static (see Hopkins, 2008; Truyts, 2010; Postlewaite, 2011, for an overview), more recently attention has turned to dynamic interdependent preferences and their interaction with social outcomes. Dioikitopoulos et al. (2019, 2020) investigated the dynamics of income and wealth inequality in an economy with dynamic positional preferences. In their model the weight on consumption relative to others in the utility function decreases as the average capital stock in the economy rises. This assumption reflects an observation that the degree of positional concern is lower in the richer countries. In a related strand of literature on aspirations, Genicot and Ray (2017) developed a theory of bidirectional interaction between individual aspirations, modelled as a reference point in the relative component of individual utility, and the distribution of income. In their model an agent derives additional utility if her bequest to her offspring exceeds an aspiration threshold which depends on the agent's own income and on the income distribution. They show that, depending on the initial aspirations, the long-run outcome can be convergence to equal distribution or divergence to income clusters.

---

[3] A cross-national survey of attitudes, beliefs and behaviour patterns conducted bianually across Europe since 2001. See https://www.europeansocialsurvey.org/.

This paper is linked to the literature on the relationship between preferences and persistence of poverty, or the so-called poverty traps. Theoretical modelling of poverty traps emerging in a long-run equilibrium started with the seminal paper by Galor and Zeira (1993), where the mechanism is driven by the fixed costs in the production technology. In the later literature the focus has shifted to preferences as the mechanism behind poverty traps.

Moav (2002) showed that non-homothetic altruistic preferences (bequests convex in income) can replace the assumption of non-convex technology. Moav and Neeman (2010, 2012) show that when individuals have positional concerns, poverty trap can be generated by the poor spending a large proportion of their income on conspicuous consumption. Borissov (2016) showed that with altruistic preferences and positional concern both in relative consumption and relative bequests those initially poor over-spend and eventually fall in a poverty trap when envy in consumption is sufficiently strong. However, envy in bequests counteracts over-spending by incentivising saving and, if it is sufficiently strong, the poverty trap can be avoided. In the literature on aspirations, an endogenous reference point in utility, determined either by costly effort by the individual (Dalton et al., 2016) or by social outcome in the macroeconomic equilibrium (Bogliacino and Ortoleva, 2014; Genicot and Ray, 2017), can lead to polarisation, with the poor trapped in a low-income cluster, if the initial distribution is sufficiently dispersed. The feature of the model behind this result is that the incentive for the poor to save and invest weakens, when their final wealth (or bequest to offspring) fall below the reference point, – the situation referred to as 'frustrated aspirations'.

In other models, over-spending and under-investment by the poor results from non-standard time preferences. Banerjee and Mullainathan (2010) and Bernheim, Ray and Yeltekin (2015) used versions of time-inconsistent preferences to show that individuals with low initial consumption, leading to higher spending on temptation goods, or with low initial assets, leading to limited self-control, can be driven into a poverty trap. In Borissov (2013) an individual time discount factor is an increasing function of income relative to the average income in the economy. Individuals with low initial income are less patient and thus spend more and invest less than those with high initial income, which leads to yet lower income and less patience in the future and, eventually, to a poverty trap.

In our model time preferences are standard, and the poor over-spend because of the envy motive to 'keep up with the Joneses', which grows in importance as inequality rises. For a given degree of inequality, the larger the negative consumption gap, the higher is the individual marginal utility of consumption. Over-spending and under-saving by the poor increase wealth inequality, leading to a higher weight of relative consumption in the utility and, thus, to a further increase in the marginal utility of consumption. This further exacerbates the incentive to spend for the poor, pushing them into a poverty trap.

Our approach is close in the spirit to that of Genicot and Ray (2018), in that the benchmark in the relative component of individual preferences is affected by social outcomes. We model positional concern as the 'keeping up with the Joneses' component in the utility of consumption. We assume that the weight on relative consumption in the utility, referred to as the strength of envy, is determined by the distributional concern: higher wealth inequality leads to stronger envy. Our model generates similar history dependence in the long-run distributional outcomes, under the assumption of the endogenous strength of the relative component, rather than endogenous reference point in Genicot and Ray (2018).

This paper is also linked to the literature on the role of intergenerational fiscal policies in the evolution of wealth inequality. Progressive taxation of wealth and, in particular, of inherited wealth, as the means of reducing inequality has been strongly advocated in this literature in the recent years (see, *inter alia*, Piketty and Saez, 2013; Saez and Zucman, 2019). Kopczuk (2015) gives an overview of the theoretical models and the empirical evidence of the redistributive role of the taxation of wealth in the form of intergenerational transfers, – in particular, the estate taxes. In much of this literature the wealth accumulation is driven by intergenerational altruism. An additional motive stemming from positional concerns was explored in Pham (2005) and Borissov and Kalk (2020).

In Pham (2005) two types of agents differ in the strength of status-seeking, and those with stronger status-seeking concern accumulate more wealth in the long run. The author analyses the relationship between inequality and growth in a setting where personal income tax is used to finance public investment in production. She shows that higher inequality caused by stronger positional concern of one type of agents can be consistent with higher growth. The reason is that larger wealth accumulated by these agents increases the total wealth and, thus, enables larger public investment.

Borissov and Kalk (2020), in an $AK$ growth model with public debt financed by distortionary labour income tax, show that a reduction in public debt can reduce inequality and increase growth in the long run. Thus, in their setting there is no trade-off between equality and growth. The evolution of inequality in Borissov and Kalk (2020) does not rely on heterogeneity in preferences or productivity of agents, who differ only in the initial endowments, and is driven entirely by the positional concern externality.

In our framework, similarly, agents are identical except for their initial endowments, and positional concern generates inequality, but, in addition, inequality affects the strength of the positional concern. The fiscal policy of taxes and public spending effectively shifts the endogenous threshold for the initial inequality. This allows, first, to put the economy on the dynamic path along which the inequality falls, and, second, after the inequality becomes sufficiently weak, to move the economy onto another dynamic path that converges to the long-run equilibrium with higher aggregate output and wealth. Thus, for any initial state such a policy eliminates trade-off between aggregate wealth and equality in wealth distribution.

## 3 The Model

The economy consists of the households who work, consume, save and leave bequests, the firms owned by households, and the government which collects taxes on labour income, capital income, and on inherited wealth, to fund public consumption.

### 3.1 Firms

The production side of the economy consists of many identical competitive firms. Every period the firms produce a homogenous good that may be consumed or invested. The production technology has constant returns to scale in two inputs, capital and labour, and so the producers can be described by a representative firm. At each time $t$, the aggregate output, $Y_t$, is determined by the Cobb-Douglas production function $Y_t = K_t^\alpha N_t^{1-\alpha}$, $0 < \alpha < 1$, where $K_t$ is the time $t$ stock of physical capital which fully depreciates during one time period, and $N_t$ is the labour input at time $t$. Factor markets are assumed to be competitive and hence the interest rate $r_t$ and the wage rate $w_t$ are equal to the marginal products of capital and labour, respectively:

$$1 + r_t = \alpha k_t^{\alpha-1}, \ w_t = (1-\alpha)k_t^\alpha, \tag{1}$$

where $k_t := K_t/N_t$ is capital per unit of labour, or the capital intensity. The output per unit of labour is $y_t := Y_t/N_t = k_t^\alpha$.

## 3.2 Government

The government collects taxes and uses the revenue to finance spending on public consumption, $G_t$, as a fixed share $\phi$ of the aggregate output:

$$G_t = \phi Y_t.$$

There is no other government spending, and the government runs balanced budget in every period. Taxes are levied on labour income and on gross capital income, comprised of wealth inherited in the form of capital and the return on capital investment. In each period $t-1$ the government announces the tax rates for period $t$, denoted by $\tau_t^w$ for labour income and $\tau_t^s$ for capital; we assume that the government can credibly commit to the next period's tax rates. Thus, the budget constraint of the government is

$$G_t = \tau_t^w w_t N_t + \tau_t^s (1+r_t) K_t = N_t \left( \tau_t^w w_t + \tau_t^s (1+r_t) k_t \right). \qquad (2)$$

Therefore, the tax rates, $\tau_t^s$ and $\tau_t^w$, satisfy

$$\alpha \tau_t^s + (1-\alpha) \tau_t^w = \phi. \qquad (3)$$

For a given $\phi$, (3) implies that $\tau_t^s$ and $\tau_t^w$ are fully determined by

$$\nu_t := \frac{1 - \tau_t^s}{1 - \tau_t^w}.$$

Indeed, we have

$$1 - \tau_t^w = \frac{1-\phi}{\alpha \nu_t + (1-\alpha)}, \quad \tau_t^w = 1 - \frac{1-\phi}{\alpha \nu_t + (1-\alpha)},$$

$$1 - \tau_t^s = \frac{\nu_t(1-\phi)}{\alpha \nu_t + (1-\alpha)}, \quad \tau_t^s = 1 - \frac{\nu_t(1-\phi)}{\alpha \nu_t + (1-\alpha)}.$$

In what follows we assume that $\phi$ is given and that the government chooses $\nu_t$.

**Assumption 0.** Parameters $\alpha$ and $\phi$ satisfy

$$\overline{\tau}_t^w := \frac{\phi}{1-\alpha} < 1, \ \overline{\tau}_t^s := \frac{\phi}{\alpha} < 1. \tag{4}$$

It is clear that if $\tau_t^w = \overline{\tau}_t^w$, then $\tau_t^s = 0$, and if $\tau_t^s = \overline{\tau}_t^s$, then $\tau_t^w = 0$. Also note that $\nu_t$ must belong to the segment $[\underline{\nu}, \overline{\nu}]$, where

$$\underline{\nu} := 1 - \overline{\tau}_t^s, \ \overline{\nu} := \frac{1}{1 - \overline{\tau}_t^w}.$$

We will refer to the set of taxes $\{\tau_t^s, \tau_t^w \mid \phi\}$ or, equivalently, $\{\nu_t \mid \phi\}$, satisfying the balanced budget condition (2)-(3), as the fiscal policy at time $t$.

## 3.3 Households

The economy is populated by successive generations of households. Time is discrete and infinite, with $t = -1, 0, 1, \ldots$. The population is constant and at any time $t$ consists of $N$ dynasties. Each individual is endowed with one unit of labour, lives for one period, and gives birth to one offspring. She receives a non-negative bequest from her parent, works, consumes, and leaves a non-negative bequest to her offspring.

Consider an individual who belongs to dynasty $j \in \{1, \ldots, N\}$ and lives in period $t$. There is no use of time other than work, and so the disposable income of this agent is $(1 - \tau_t^s)(1+r_t)s_{t-1}^j + (1 - \tau_t^w) w_t$, where $s_{t-1}^j \geq 0$ is the bequest, in the form of capital investment, left by her parent in the previous period, $r_t$ is the net return on investment, or the interest rate, $w_t$ is labour income that is equal to the wage rate, $\tau_t^s$ is the rate of tax on the inherited wealth and capital income, and $\tau_t^w$ is the rate of tax on labour income. She divides her disposable income between her personal consumption, $c \geq 0$, and a bequest she leaves to her offspring, $s \geq 0$.[4] Thus, her budget constraint is $c + s = (1 - \tau_t^s)(1 + r_t)s_{t-1}^j + (1 - \tau_t^w) w_t$.

Individual preferences are represented by the following utility function:

$$
\begin{aligned}
u_t(c, s) &= \ln(c + \gamma_t [c - \bar{c}_t]) \\
&\quad + \delta \ln((1 - \tau_{t+1}^w) w_{t+1} + (1 - \tau_{t+1}^s)[1 + r_{t+1}] s),
\end{aligned}
\tag{5}
$$

where

$$\bar{c}_t := \frac{1}{N} \sum_{j=1}^{N} c_t^j$$

This utility function describes envy and altruism. Specifically, an individual in period $t$ compares her consumption level $c$ with a reference level of consumption, assumed to be equal to the average level $\bar{c}_t$ of consumption of generation $t$. The higher (lower) is the individual consumption relative to the reference level, the higher (lower) is the utility of consumption. The weight on the relative component, $\gamma_t \geq 0$, common for all agents at time $t$, measures the extent of consumption-related positional concerns, or the degree of envy; the value of zero means no positional concerns. The agent also derives utility from the disposable income of her heir, with $\delta > 0$ measuring the degree of parental altruism.

We do not include positional concerns in this last component of the utility function. Although there is some documented evidence of positionality in parents' attitude to children's intelligence and education (Celse, 2012), bequests have low visibility and for that reason play little role in social perceptions and comparisons (Heffetz and Frank, 2011; Alvarez-Cuadrado and

---

[4]To simplify the presentation, we assume that the only motive for saving is bequests.

Long, 2012).[5]

Each individual chooses consumption and bequest so as to maximise their utility subject to the budget constraint:

$$\max_{c\geq 0, s\geq 0}\{\ln\left(c + \gamma_t \left[c - \bar{c}_t\right]\right) + \delta \ln((1 - \tau^w_{t+1}) w_{t+1} + (1 - \tau^s_{t+1}) [1 + r_{t+1}] s)\}$$
$$\text{subject to } c + s = \left[(1 - \tau^w_t) w_t + (1 - \tau^s_t)(1 + r_t) s^j_{t-1}\right]. \tag{6}$$

We assume that each individual ignores the effect of her consumption on average consumption and the effect of her savings on the aggregate capital stock. (Equivalently, one can assume that the economy is populated by $N$ types of households, where each type consists of a continuum of identical atomless households.)

Our main assumption is that the degree of envy depends on the distribution of the inherited wealth,

$$\gamma_t = \gamma(s^1_{t-1}, ..., s^N_{t-1}) \tag{7}$$

and is increasing in wealth inequality, as stated below.

**Assumption 1.** $\gamma(\cdot)$ is a symmetric (anonymous) continuous 0-homogeneous function defined on $\mathbb{R}_+ \setminus \{0\}$ such that $\gamma\left(s^1, ..., s^N\right) > \gamma\left(s'^1, ..., s'^N\right)$ whenever $(s^1, ..., s^N)$ and $(s'^1, ..., s'^N)$

---
[5]Borissov (2016) analyses a model with a similar utility function but where fixed positional concerns are present in both components, and are weaker in the parental altruism component than in the own consumption component.

satisfy

$$(i) \quad s^1 \leq ... \leq s^N;\ s'^1 \leq ... \leq s'^N \tag{8}$$

$$(ii) \quad \frac{\sum_{i=1}^{M} s^i}{\sum_{j=1}^{N} s^j} \leq \frac{\sum_{i=1}^{M} s'^i}{\sum_{j=1}^{N} s'^j},\ \forall M = 1, ..., N, \tag{9}$$

$$(iii) \quad \exists \overline{M} \in \{1, ..., N\} : \frac{\sum_{i=1}^{\overline{M}} s^i}{\sum_{j=1}^{N} s^j} < \frac{\sum_{i=1}^{\overline{M}} s'^i}{\sum_{j=1}^{N} s'^j}. \tag{10}$$

Condition (8) is just an anonymous ordering. Conditions (9) and (10) describe first-order stochastic dominance of distribution $(s^j)_{j=1}^N$ over distribution $(s'^j)_{j=1}^N$: the proportion of wealth held by $M$ poorest households under distribution $(s^j)_{j=1}^N$ is no greater than that under distribution $(s'^j)_{j=1}^N$ for every $M \leq N$ and is strictly less for a least one $M$. Note that Assumption 1 does not require $\sum_{j=1}^{N} s^j = \sum_{j=1}^{N} s'^j$ and so applies to comparisons when the levels of aggregate wealth are different.

We postulate the dependence of positional concern on inequality very broadly, stating it as the ranking of cumulative distributions of inherited wealth, without assuming a parametrised functional form. Assumption 1 is consistent, for example with assuming that $\gamma_t$ is increasing in the Gini coefficient. While the utility of an individual agent depends on inequality, Assumption 1 does not describe inequality aversion (or, indeed, inequality-loving). Because consumption levels are determined endogenously, an increase in the inequality of wealth distribution, in general, may change the configuration of consumption in a way that will not necessarily lead to utility loss for every agent even though the degree of envy increases for everyone.

## 4 Equilibria

We now proceed to defining the market equilibrium in this economy.

**Definition 1** *Let the fiscal policies at times $t$ and $t+1$, $\{\nu_t, \nu_{t+1}; \phi\}$, be given. Let the bequests $\{s_{t-1}^j \geq 0, \ j = 1, \ldots, N\}$, left by the agents in period $t-1$ also be given, and let $k_t = \frac{\sum_{j=1}^N s_{t-1}^j}{N} > 0$. A tuple $\{(c_t^j, s_t^j)_{j=1}^N, k_{t+1} | \{\nu_t, \nu_{t+1}; \phi\}\}$ constitutes a time $t$ temporary equilibrium if
i) for $\gamma_t = \gamma(s_{t-1}^1, \ldots, s_{t-1}^N)$, $(c_t^j, s_t^j)$ solves (6) at $w_t$ and $1 + r_t$ given by (1),
ii) $k_{t+1} = \frac{\sum_{j=1}^N s_t^j}{N} > 0$, and iii) $N_t = N$ for every $t$.*

Let
$$\xi := \frac{1-\alpha}{\alpha}.$$

For all $t = 0, 1, \ldots$, we have $w_t = \xi(1+r_t)k_t$, and hence

$$\ln([1-\tau_{t+1}^w]w_{t+1} + [1-\tau_{t+1}^s](1+r_{t+1})s)$$
$$= \ln([1-\tau_{t+1}^s][1+r_{t+1}]) + \ln(\frac{\xi}{\nu_{t+1}}k_{t+1} + s).$$

Therefore, the optimisation problem of household $j$ in period $t$ can be formulated as

$$\begin{cases} \max_{c \geq 0, s \geq 0}\{\ln(c + \gamma_t[c - \bar{c}_t]) + \delta \ln(\frac{\xi}{\nu_{t+1}}k_{t+1} + s)\} \\ \text{subject to } c + s = (1-\tau_t^s)(1+r_t)(\frac{\xi}{\nu_t}k_t + s_{t-1}^j) \end{cases}. \quad (11)$$

### 4.1 Existence and uniqueness

To guarantee the existence of a time $t$ equilibrium it is necessary for (11) to have feasible $c$ and $s$ such that the consumption level $c$ is higher than the reference point, i.e. it is necessary for the inequality $(1 - \tau_t^s)(1 + r_t)(\frac{\xi}{\nu_t}k_t + s_{t-1}^j) > \frac{\gamma_t}{1+\gamma_t}\bar{c}_t$ to hold. To ensure this inequality, we assume that

$$\max\left\{\gamma(s^1, ..., s^N) \Big| \sum_{j=1}^{N} s^j = 1,\ s^j \geq 0,\ j = 1, ..., N\right\} < \widehat{\gamma}(\bar{\nu}),$$

where

$$\widehat{\gamma}(\nu) := \frac{\frac{\xi}{\nu}\left[\frac{\xi}{\nu} + 1 + \delta\right]}{\frac{\xi}{\nu} + 1}.$$

Note that

$$\widehat{\gamma}(\bar{\nu}) := \min_{\nu \in [\underline{\nu}, \bar{\nu}]} \widehat{\gamma}(\nu).$$

**Proposition 1** *For a given set of fiscal policies at times $t$ and $t + 1$, $\{\nu_t, \nu_{t+1}; \phi\}$, and any $\{(s_{t-1}^j)_{j=1}^N, k_t\}$ such that $s_{t-1}^j \geq 0$, $j = 1, \ldots, N$, and $k_t = \frac{\sum_{j=1}^{N} s_{t-1}^j}{N} > 0$, there exists a unique time $t$ temporary equilibrium $\{(c_t^j, s_t^j)_{j=1}^N, k_{t+1} | \{\nu_t, \nu_{t+1}; \phi\}\}$.*

See the Appendix for all proofs.

Next, to describe the evolution of our dynamic economy, we define an equilibrium path as a sequence of time $t$ equilibria.

**Definition 2** *Let the bequests $\{s_{-1}^j \geq 0,\ j = 1, \ldots, N\}$, left by the agents who live in period $t = -1$ be given. Let further $k_0 = \sum_{j=1}^{N} s_{-1}^j > 0$. A sequence $\{(c_t^j, s_t^j)_{j=1}^N, k_{t+1} | \{\nu_t, \nu_{t+1}; \phi\}\}_{t=0}^{\infty}$ constitutes an equilibrium path starting from $(s_{-1}^j)_{j=1}^N$ if for each $t = 0, 1, \ldots$, $\{(c_t^j, s_t^j)_{j=1}^N, k_{t+1} | \{\nu_t, \nu_{t+1}; \phi\}\}$ is a time $t$ temporary equilibrium.*

**Remark 1** *It should be emphasized that the formal definition of an intertemporal equilibrium in a more general setting requires that the tax rates are given* *for all $t = 0, 1, ...$. In our model, to construct an initial segment of an intertemporal equilibrium up to time $T$, we need to know the tax rates only up to time $T + 1$. This is because to construct a temporarily time t equilibrium, it is sufficient to know the state of the economy at time t and the tax rate at time $t+1$. This simplification is possible because of the Cobb-Douglas form of the production function and the log-linear form of the utility function. Therefore, in this setting we can assume that the government needs to announce the tax rates only one period ahead.*

The following existence and uniqueness result follows directly from Proposition 1.

**Theorem 1** *Let the sequence of fiscal policies for all $t = 0, 1, ...,$ $\{\nu_0, \nu_1, \ldots; \phi\}$, be given. For any $(s_{-1}^j)_{j=1}^N$ such that $\sum_{j=1}^N s_{t-1}^j > 0$, there exists a unique equilibrium path $\{(c_t^j, s_t^j)_{j=1}^N, k_{t+1} \,|\, \{\nu_0, \nu_1, \ldots; \phi\}\}_{t=0}^\infty$ starting from $(s_{-1}^j)_{j=1}^N$.*

## 4.2 Steady state

Now, to characterise the distributional properties of an economy in the long run we focus on a fiscal policy that is constant over time. Formally, suppose that $\nu \in [\underline{\nu}, \bar{\nu}]$ is given and that the tax rates are constant:

$$\tau_t^w = \tau^w = 1 - \frac{1-\phi}{\alpha\nu + (1-\alpha)}, \tag{12}$$

and

$$\tau_t^s = \tau^s = 1 - \frac{\nu(1-\phi)}{\alpha\nu + (1-\alpha)} \tag{13}$$

for $t = 0, 1, ....$

**Definition 3** *A tuple $\{(c^j, s^j)_{j=1}^N, k \,|\, \{\nu, \phi\}\}$ is a steady-state equilibrium with fiscal policy $\{\nu, \phi\}$ if $k > 0$ and the sequence $\{(c_t^j, s_t^j)_{j=1}^N, k_{t+1} \,|\, \{\nu, \phi\}\}_{t=0}^\infty$ given by*

$$k_{t+1} = k; \ (c_t^j, s_t^j) = (c^j, s^j), \ j = 1, \ldots, N, \ \forall t = 0, 1, 2, \ldots$$

*is an equilibrium path starting from $(s^j)_{j=1}^N$.*

We will now show that our economy can exhibit two types of steady-state equilibria: a polarised, or a two-class equilibrium, and an egalitarian equilibrium. In a *polarised equilibrium* the population is divided into two classes, the rich and the poor; only the rich leave positive bequests. Thus, an individual born into a poor household starts with zero initial wealth and, in turn, leave nothing to her offspring. In an *egalitarian equilibrium* all dynasties have the same consumption levels and leave the same positive bequests (and, therefore, are rich), so that all individuals in a newly born cohort have the same initial wealth.

**Definition 4** *A steady-state equilibrium $\{(c^j, s^j)_{j=1}^N, k \,|\, \{\nu, \phi\}\}$ is egalitarian if*

$$s^j = k \ (\text{and hence } c^j = (1-\phi)k^\alpha - k) \ j = 1, \ldots, N. \quad (14)$$

Let

$$s(\gamma, m, \nu) := \frac{1-\phi}{\alpha + \frac{1}{\nu}(1-\alpha)} \frac{\alpha \delta \left[1 + m\frac{\xi}{\nu} + \gamma(1-m)\right]}{\left[1 + \delta + m\frac{\xi}{\nu}\right][1+\gamma] - m\delta\gamma}. \quad (15)$$

Note that $s(\gamma, m, \nu)$ is decreasing in $\gamma$ for any $\nu$ and $m \in (0, 1]$.

Let $k(\gamma, m, \nu)$ be defined as the positive solution to the following equation in $k$:

$$k = s(\gamma, m, \nu)k^\alpha.$$

We show below (Propositions 2 and 3) that, in a steady-state equilibrium with fiscal policy $\{\tau_s, \tau_w, \phi\}$, the economy's savings rate is equal to $s(\gamma, m, \nu)$ and the capital intensity is equal to $k(\gamma, m, \nu)$, where $\gamma$ is the degree of envy and $m$ is the population share of the rich.

Let $\gamma^*(\nu)$ denote the solution of the following equation:

$$s(\gamma, m, \nu) = \frac{1-\phi}{\alpha + \frac{1}{\nu}(1-\alpha)} \frac{\alpha\delta}{1+\delta}.$$

It is straightforward to show that it is given by

$$\gamma^*(\nu) = \frac{\delta\xi}{\xi + \nu}, \tag{16}$$

and that it is a decreasing function of $\nu$,

$$\frac{d\gamma^*(\nu)}{d\nu} < 0, \gamma^*(\nu) < \widehat{\gamma}(\nu) \text{ for } \nu \in [\underline{\nu}, \bar{\nu}]$$

Note that for any $m \in (0, 1]$,

$$s(\gamma, m, \nu) \gtreqless \frac{1-\phi}{\alpha + \frac{1}{\nu}(1-\alpha)} \frac{\alpha\delta}{1+\delta} \Leftrightarrow \gamma^*(\nu) \gtreqless \gamma.$$

Further, note that for a given $\phi > 0$

$$\frac{\partial s(\gamma, m, \nu)}{\partial \nu} > 0 \ \forall m \in (0, 1] \ \forall \gamma$$

In other words, the savings rate is increasing in $\tau_w$ or, equivalently, decreasing in $\tau_s$.

Let

$$\Gamma_n := \gamma(\underbrace{0, ..., 0}_{N-n}, \underbrace{\frac{1}{n}, ..., \frac{1}{n}}_{n}).$$

Using (7), one can see that $\Gamma_1 > \Gamma_2 > ... > \Gamma_N$.

The next two propositions describe the properties of the egalitarian and the polarised equilibria.

**Proposition 2** *There is a unique egalitarian steady-state equilibrium $\{(c^j, s^j)_{j=1}^N, k \,|\, \{\nu, \phi\}\}$ described by $k = k(\Gamma_N, 1)$ and (14).*

**Proposition 3** *If a non-empty subset $J$ of the set of dynasties with cardinality $|J|$ is such that $\Gamma_{|J|} > \gamma^*(\nu)$, then there exists a steady-state equilibrium $\{(c^j, s^j)_{j=1}^N, k \,|\, \{\nu, \phi\}\}$ such that*

$$s^j > 0, \ j \in J; \ s^j = 0, \ j \notin J.$$

*In this equilibrium*
$$k = k(\Gamma_{|J|}, \frac{|J|}{N}, \nu);$$
$$s^j = \frac{N}{|J|} k \text{ and } c^j = \frac{N}{|J|} [\phi k^\alpha - k] - (\frac{N-|J|}{|J|})(1-\alpha)(1-\tau^w) k^\alpha, \ j \in J;$$
$$s^j = 0 \text{ and } c^j = (1-\alpha)(1-\tau^w) k^\alpha, \ j \notin J.$$

An equilibrium described in Proposition 3 is a polarised steady-state equilibrium whenever $|J| < N$.

## 4.3 Equilibrium Dynamics of Wealth Distribution

Having established the existence of two types of steady-state equilibria, we now investigate how consumption, bequests, and the capital stock in this economy change along the equilibrium path when the tax rates are constant. It turns out that in the long run the economy converges either to the egalitarian or to a polarised steady-state equilibrium, depending on the initial distribution of wealth, $(s_{-1}^j)_{j=1}^N$.

Let $\{(c_t^j, s_t^j)_{j=1}^N, k_{t+1} \,|\, \{\nu, \phi\}\}_{t=0}^\infty$ be an equilibrium path starting from $(s_{-1}^j)_{j=1}^N$ such that $\sum_{j=1}^N s_{-1}^j > 0$. Without loss of generality we assume $s^1_{-1} \geq s^2_{-1} \geq \ldots \geq s^N_{-1}$. By $L$ we denote the number of agents $j$ such that $s^j_{-1} = s^1_{-1}$:

$$s^1_{-1} = s^2_{-1} = \ldots = s^L_{-1} > s^{L+1}_{-1} \geq \ldots \geq s^N_{-1}.$$

That is, $L \geq 1$ is the number of agents with the highest initial endowment of wealth.

It is not difficult to check that if $s^j_{t-1} \geq s^i_{t-1}$, then $s^j_t \geq s^i_t$. Moreover, if $s^j_{t-1} > s^i_{t-1}$ and $s^j_t > 0$, then $s^j_t > s^i_t$. Therefore,

$$s^1_t = s^2_t = \ldots = s^L_t > s^{L+1}_t \geq \ldots \geq s^N_t, \ t = 0, 1, \ldots. \tag{17}$$

In other words, the top $L$ equally rich heirs, in turn, leave the largest (equal) bequests to their offspring.

Let $M(t)$ be the number of agents who leave positive bequests in period $t$:

$$s^j_t > 0, \ j = 1, \ldots, M(t); \ s^j_t = 0, \ j = M(t) + 1, \ldots, N.$$

**Theorem 2** *1) If $\gamma(s^1_{-1}, ..., s^N_{-1}) < \gamma^*(\nu)$, then:*

$$M(t) = N, \ t = 0, 1, \ldots, \tag{18}$$

$$k_{t+1} = s(\gamma_t, 1, \nu)k_t^\alpha, \ t = 0, 1, \ldots, \tag{19}$$

$$\lim_{t \to \infty} \frac{s^j_{t-1}}{k_t} = 1, \ j = 1, \ldots, N, \tag{20}$$

*and hence*

$$\lim_{t \to \infty} \gamma_t = \Gamma_N$$

*and*

$$\lim_{t \to \infty} k_t = k(\Gamma_N, 1, \nu). \tag{21}$$

2) If $\gamma(s_{-1}^1, ..., s_{-1}^N) > \gamma^*(\nu)$, then the sequence $\{M(t)\}_{t=0}^{\infty}$ is non-increasing and there exists $T$ such that for $t = T+1, T+2, \ldots$,

$$M(t) = L, \tag{22}$$

$$k_{t+1} = s(\Gamma_L, L/N, \nu)k_t^\alpha, \tag{23}$$

$$\frac{s_{t-1}^j}{k_t} = \frac{N}{L}, \ j = 1, \ldots, L; \ s_{t-1}^j = 0, \ j = L+1, \ldots, N; \tag{24}$$

and

$$\lim_{t \to \infty} k_t = k(\Gamma_L, L/N, \nu). \tag{25}$$

**Remark 2** *It is easily checked that the long-run capital stock and, hence, the long-run output in the egalitarian steady-state equilibrium is higher than in any polarised equilibrium. This is true for any values of the initial capital stock. Consider two countries, say, A and B, with the same fiscal policies (the same $\phi$ and $\nu$), and country A initially richer (with higher initial capital stock) than country B. Suppose that country A is on the equilibrium path to the polarised steady-state equlibrium because of a high initial inequality, whereas country B is on the equilibrium path to the egalitarian steady-state equilibrium. Then the initially poorer country B will gradually overtake the initally richer country A in the capital stock and output. When the fiscal policies are different, the situation is more delicate and interesting. This is discussed below in Section 5.2.*

## 5 Discussion

The model predicts that in the long run the distributional and aggregate properties of the equilibrium depend on the initial conditions and on the choice of the combination of taxes. In what follows we discuss some policy implications of these results.

## 5.1 Path dependence

Theorem 2 tells us about the importance of the initial conditions for the long-run outcome in the economy, or the path-dependence. For a given fiscal policy, if the initial envy among households is strong, then the economy eventually becomes poor and polarises into two classes, the rich and the poor. If the initial envy is weak, the economy eventually becomes rich and egalitarian.

Since the strength of envy is determined by wealth distribution, for a given fiscal policy, the outcome depends both on the degree of inequality and on the sensitivity of the degree of envy to inequality. Thus, an egalitarian outcome is more likely, the less unequal is the initial distribution of wealth, or the less sensitive is the degree of envy to inequality, or both. Put differently, from the same initial wealth distribution inequality can increase over time in one society and fall in another society if in the first society people care more about consumption relative to others that in the second society.

Finally, using (16), it is easy to check that, given the set of taxes, the envy threshold, $\gamma^*(\nu)$, is higher, the stronger is the degree of parental altruism. Higher weight put on the offspring's wealth counteracts the tendency to over-consume caused by envy motive. Thus, the model predicts that an egalitarian outcome is more likely, the higher is the parents' concern about their children.

Theorem 2 also implies that an egalitarian equilibrium in the long run can be achieved by a one-off policy intervention in the form of redistribution of wealth without changing fiscal policy. It is sufficient to redistribute inherited wealth once in any time period so as to bring $\gamma_t$ below the threshold $\gamma^*(\nu)$, – the internal dynamic forces will lead the economy to the egalitarian outcome.

## 5.2 The role of fiscal policy

The egalitarian outcome in the long run can also be achieved by a fiscal policy reform. Consider the choice of tax rates, $\tau^w$ and $\tau^s$ (in our framework this is reduced to the choice of $\nu$ for a given ratio $\phi$ of public consumption to output) having in mind two objectives: a higher aggregate wealth (or, equivalently, output) and a lower level of income inequality. Common wisdom suggests that there is a trade-off between these objectives. If the goverment wants to stimulate output, it should set the tax on the capital as small as possible. This, however, for a fixed $\phi$ will necessitate an increase the labour income tax and, hence, exacerbate income inequality. Conversely, if the government aims to reduce inequality, it should decrease the labour income tax and increase the tax on capital. This, however, will undermine the incentives to save and result in lower output. For polarised equilibria in our model this logic works as well, but only in the short run. In the long run, the picture is more subtle.

Suppose that the initial state is characterised by some $(s^j_{-1})^N_{j=1}$, so that the degree of envy at time $t = 0$ is equal to

$$\gamma_0 = \gamma(s^1_{-1}, ..., s^N_{-1}).$$

The government needs to maintain $\phi$ at a given level and can do this by using any combination of taxes that ensures balanced budget in every period, i.e. it has a free choice of $\nu$. The long-run dynamics of the economy crucially depends on $\nu$ because the degree of envy, $\gamma(\nu)$, is decreasing in $\nu$.

First, consider the case where $\tau_w$ is high and $\tau_s$ is low and hence $\nu = \frac{1-\tau_s}{1-\tau_w}$ is sufficiently high. Suppose that the tax rates, and, therefore, $\nu$, remain constant. Namely, let $\nu = \nu'$, where $\nu'$ is such that

$$\gamma_0 = \gamma(s^1_{-1}, ..., s^N_{-1}) > \gamma^*(\nu').$$

In this case the dynamics of the economy is fully described by the second part of Theorem 2: $(i)$ the the population divides into two classes, $(ii)$ $\gamma_t$ converges to $\Gamma_L$, where $L$ is the number of the initially richest individuals, $(iii)$ the savings rate converges to $s(\Gamma_L, L/N, \nu')$, and $(iv)$ the *per capita* capital stock converges to $k(\Gamma_L, L/N, \nu')$. On Fig. 1, the polarised steady-state equilibrium to which the economy converges is presented by point $E_1$.

Now consider the case where $\tau_w$ is lower and, therefore, $\tau_s$ is higher. This means that $\nu = \frac{1-\tau_s}{1-\tau_w}$ is lower than in the previous case. Namely, let in this case $\nu = \nu''$, where $\nu''$ is such that

$$\gamma_0 = \gamma(s^1_{-1}, ..., s^N_{-1}) < \gamma^*(\nu'').$$

In this second case the scenario is as described in the first part of Theorem 2: $(i)$ the economy settles on the egalitarian regime, $(ii)$ $\gamma_t$ converges to $\Gamma_N$, $(iii)$ the savings rate converges to $s(\Gamma_N, 1, \nu'')$, and $(iv)$ the *per capita* capital stock converges to $k(\Gamma_N, 1, \nu'')$. On Fig. 1, the egalitarian steady-state equilibrium to which the economy converges in the second case is presented by point $E_2$.

[FIGURE 1 ABOUT HERE]

[Fig. 1: Polarised ($E_1$) and egalitarian ($E_2$ and $E_3$) equilibria.]

The relationship between the savings rates in the two steady-state equilibria depends on the initial degree of envy $\gamma_0$, determined by the initial level of inequality, and on the values of $\nu'$ and $\nu''$. In particular, it can happen that the savings rate in the polarised equilibrium is higher than that in the egalitarian equilibrium, $s(\Gamma_L, L/N, \nu') > s(\Gamma_N, 1, \nu'')$, as in Fig. 1. In this case, there is a long-run trade off between equality and output.

To illustrate this trade-off, suppose that $L$ is given, and denote by $S(\gamma_0, \nu)$ the savings rate in the equilibrium to which the economy converges depending on $\gamma_0$ and $\nu$. Figures 2(a)–(c) illustrate this dependence. In this figures, $\nu(\gamma_0)$ is the solution to the following equation in $\nu$: $\gamma^*(\nu) = \gamma_0$. Clearly, $\nu(\gamma_0)$ is increasing in $\gamma_0$. The three different values of $\gamma_0$ presented in this figure, $\gamma_0'$, $\gamma_0''$, $\gamma_0'''$, are such that $\gamma_0' > \gamma_0'' > \gamma_0''' = \Gamma_N$, and, hence, $\nu(\gamma_0') > \nu(\gamma_0'') > \nu(\gamma_0''')$. In other words, Fig. 2(a) corresponds to the case where the initial inequality is high, Fig. 2(b) to the case of intermediate initial inequality and Fig. 2(c) to an egalitarian initial wealth distribution[6].

[FIGURE 2(a) ABOUT HERE]

[Fig. 2(a): Long-run savings rate with high initial inequality.]

[FIGURE 2(b) ABOUT HERE]

[Fig. 2(b): Long-run savings rate with intermediate initial inequality.]

[FIGURE 4 ABOUT HERE]

[Fig. 2(c): Long-run savings rate with egalitarian initial distribution.]

We can see that, if the initial distribution of wealth is egalitarian, then there is no long-run trade off between equality and aggregate wealth. The tax rates that lead to the highest savings rate in the long run ($\tau_w = \overline{\tau}_w$ and $\tau_s = 0$) will not push the economy out of the egalitarian regime.

---

[6] Here we assume that $\Gamma_N < \gamma^*(\bar{\nu})$.

If, on the contrary, the initial wealth inequality is high, then a high labour income tax and a low capital tax will lead the economy to a polarised steady-state equilibrium. Moreover, it is easy to show that in this steady-state equilibrium, an increase in the labour income tax will result in a higher level of income inequality. On the other hand, implementing a combination of tax rates that would lead the economy from this initial state to the egalitarian steady-state equilibrium, would result in a lower savings rate and, consequently, in a lower aggregate output. Thus, in the case of a high level of initial inequality, the trade-off between equality and output exists.

However, this is not the end of the story because the above argument holds true only under the assumption that the tax rates are set once and for ever. If the government can change the tax rates in some period and keep them constant thereafter, the trade-off between equality and output in the long run can be overcome completely or at least to some extent. Indeed, if the wealth inequality is high, then the following scenario is possible. At the initial stage, the government sets the tax rates in such a way that the economy starts moving towards the egalitarian equilibrium (for example, toward point $E_2$ on Fig. 1). Eventually, when the level of wealth inequality becomes sufficiently low, the government implements a tax reform, by lowering the capital tax and increasing the labour income tax, while maintaining its spending-to-output ratio and balanced budget. The reform moves the economy onto a different path that converges to another egalitarian steady state equilibrium,[7] with a higher savings rate and, therefore, with a higher output and aggregate wealth (for example, point $E_3$ on Fig. 1).

---

[7]Such a reform is consistent with the definition of the equilibrium path in our model,

# 6 Conclusion

In this paper we analysed the dynamics of wealth inequality in an economy where agents' preferences are characterised by envy towards their peers and altruism towards their offsprings. The novel feature is the dynamic interaction between envy and inequality: we assume that higher inequality leads to stronger envy. Specifically, higher inequality in the inherited wealth increases the strength of the positional concern.

Our model generates history dependence of the long-run outcomes. It predicts that the distribution of wealth among households can become less disperse, or less unequal over time if the initial inequality is not too large, or the initial degree of envy is not too strong. Otherwise, the distribution of wealth eventually becomes more unequal. The initial conditions determine whether in the long run the economy becomes rich and egalitarian or poor and polarised. Importantly, this result implies that a one-off redistribution can reverse the negative trend and eventually lead to a higher and more evenly distributed aggregate wealth. As the economy moves towards the egalitarian equilibrium, the inequality reduces and the output grows, along with the capital stock. This suggests negative endogenous relationship between wealth inequality and aggregate wealth. While the theoretical and empirical literature on the relationship between wealth inequality and aggre-

---

for the reason explained in Remark 1.

gate wealth is immense, with researchers looking at a wide range of measures and factors,[8] our model offers a novel channel linking endogenous individual preferences and macroeconomic performance.

We also show that a balanced-budget fiscal policy of public spending financed by taxes on capital and labour income can be used to overcome this history dependence and lead the economy to the egalitarian equilibrium for any degree of initial inequality. First, a combination of low labour income tax and high capital tax helps reduce inequality along the equilibrium path. While a high capital tax results in lower savings and lower output in the long run, this can be remedied by a one-off fiscal reform. When the inequality falls below the threshold, a reduction in capital tax compensated by an increase in the labour income tax can be constructed in such a way that the economy continues converging to the egalitarian long-run equilibrium. An important feature of the model is that the egalitarian equilibrium is characterised by higher wealth than any polarised equilibrium. Hence, in contrast to the standard result, there is no trade-off between equality and aggregate wealth.

This framework can be applied also to the analysis of the distribution of wealth across countries in the context of globalisation and international capital flows. Suppose, an economy in our model is an economic union, with an individual agent now representing a member country, and with free capital and labour flows between countries, such as in the European Union (EU). Then, according to our model predictions, richer countries will command higher proportion of total capital than poorer countries. Empirically, this will be reflected in capital ownership: citizens of rich countries will own much of capital in the poor countries. Indeed, among the EU countries, the Gross National Income (GNI) is below the Gross Domestic Product (GDP) for mostly poorer EU member countries (such as Czech Republic and Romania), while for mostly rich ones (France and Germany) the GNI is above the GDP. The ratio of GNI to GDP is positively correlated to both the per capita GNI and per capita GDP, with the sample correlation coefficients of 0.57 and 0.54, respectively.[9]

---

[8]See, for example, Piketty and Zucman (2015) for a survey.

Assuming that each dynasty in our model represents a country, our model predicts that globalisation with free flows of capital and labour may eventually lead to divergence and polarisation of countries if the initial inequality among them is sufficiently high. This trend, however, can be reversed, and wealth convergence and economic growth can be achieved in the long run, by redistribution of capital among countries. This result links our paper to the models of international capital flows which assume cross-country heterogeneity in endowments or preferences. When countries differ in time preferences, capital flows from more patient into less patient countries. Vidal (2000) analysed the distributional and welfare implications of capital flows generated by the difference in strength of parental altruism between countries in an overlapping generations (OLG) framework. In Vidal's model the altruism is non-paternalistic, or pure, so that the strength of altruism towards children is, in effect, an intergenerational discount factor. In our model heterogeneity is in the initial endowments, leading to the differences in the positional

---

[9]These correlation coefficients were computed for the cross-section of the EU in total and 26 EU countries in 2017. Ireland and Luxembourg were excluded as strong outliers, owing to their tax haven status. Source: Eurostat.

concern in preferences that have distributional consequences.

Our model is not intended to explain in its entirety the existing inequality within and across countries or, indeed, to offer an ultimate remedy. Instead, the model suggests that interaction between positional and distributional concerns in preferences is a mechanism that could be contributing to the dynamics of inequality, generating the outcomes consistent with certain empirical observations. Verifying causal link from inequality to the strength of positional concerns and quantifying such an effect is a challenging task. Yet, increasing availability of longitudinal data, including life satisfaction surveys, and growing interest to the dynamics of inequality and social comparisons might give rise to the empirical investigation of this mechanism.

A stylised and analytically tractable approach to modelling dynamic endogenous positional preferences developed in this paper can be used for the analysis of altlernative redistributive fiscal policies, such as the tax and transfer systems, provision of public goods, or productive government spending financed by tax or public debt. The analysis can be applied to internal policies in a single country or to the international policies in an economic union or in a globalised world economy. These extensions are left for the future research.

# Appendix

**Proof of Proposition 1.** It follows from Proposition 1 in Borissov (2016). □

To prove Proposition 2, Proposition 3 and Theorem 2 we make several

preliminary comments. Let

$$z_t := \frac{\gamma_t}{1+\gamma_t}.$$

Following Borissov (2016), one can show that if $z_t < \frac{\widehat{\gamma}(\bar{\nu})}{1+\widehat{\gamma}(\bar{\nu})}$, then in a time $t$ temporary equilibrium we have $z_t \bar{c}_t < (1-\tau^s)(1+r_t)(\frac{\xi}{\nu_t}k_t + s_{t-1}^j)$. This inequality implies that there is a unique solution to problem (11), $(c_t^j, s_t^j)$, which is determined as follows:

$$s_t^j = \max\{0, \frac{\delta(1-\tau_t^s)(1+r_t)(\frac{\xi}{\nu_t}k_t + s_{t-1}^j) - (\delta z_t \bar{c}_t + \frac{\xi}{\nu_{t+1}}k_{t+1})}{1+\delta}\}, \quad (26)$$

$$c_t^j = (1-\tau_t^s)(1+r_t)(\frac{\xi}{\nu_t}k_t + s_{t-1}^j) - s_t^j.$$

Obviously, in a temporary time $t$ equilibrium $\{(c_t^j, s_t^j)_{j=1}^N, k_{t+1} | \{\nu_t, \nu_{t+1}; \phi\}\}$, we have

$$\bar{c}_t + k_{t+1} = (1-\tau_t^s)(1+r_t)(\frac{\xi}{\nu_t}+1)k_t$$

and hence

$$\bar{c}_t = (1-\tau_t^s)(1+r_t)(\frac{\xi}{\nu_t}+1)k_t - k_{t+1}. \quad (27)$$

For the results stated in Proposition 2, Proposition 3 and Theorem 2 we restrict our attention to the equilibria with constant fiscal policies, $\nu_t = \nu$. It is easily checked that for any $x \in (0, 1]$,

$$\gamma \gtreqless \gamma^*(\nu) \Leftrightarrow \frac{1-\phi}{\alpha+\frac{1}{\nu}(1-\alpha)}\frac{\alpha\delta}{(1+\delta)s(\gamma, x, \nu)} \gtreqless 1 \quad (28)$$

$$\Leftrightarrow \alpha\delta\frac{\xi}{\nu} - \frac{\alpha\delta\gamma}{1+\gamma}\frac{\nu(1-\phi)}{\alpha\nu+(1-\alpha)}(\frac{\xi}{\nu}+1) + \left[\frac{\delta\gamma}{1+\gamma} - \frac{\xi}{\nu}\right]s(\gamma, 1, \nu) \lesseqgtr 0$$

For what follows we need three claims for which the proofs are similar to those in Borissov (2016).

**Claim 1** $M(t) \geq M(t-1) \Rightarrow k_{t+1} = \frac{s(\gamma_t, M(t)/N, \nu)}{\alpha}(1+r_t)k_t$.

**Claim 2** $\gamma_t < \gamma^*(\nu) \Rightarrow M(t) = N$.

**Claim 3** $\gamma_t > \gamma^*(\nu) \Rightarrow M(t) \leq M(t-1)$.

**Lemma 1** *If $\gamma_t < \gamma^*(\nu)$, then*

$$s_{t-1}^j > s_{t-1}^i \Rightarrow \frac{s_t^i}{s_t^j} > \frac{s_{t-1}^i}{s_{t-1}^j}, \tag{29}$$

*and if $\gamma_t > \gamma^*(\nu)$, then*

$$s_{t-1}^j > s_{t-1}^i > 0 \ \& \ s_t^j > 0 \Rightarrow \frac{s_t^i}{s_t^j} < \frac{s_{t-1}^i}{s_{t-1}^j}. \tag{30}$$

**Proof.** 1) Let $\gamma_t < \gamma^*(\nu)$. Taking account of Claims 1 and 2, (13), (26), and (27), for all $n = 1, ..., N$, we have

$$\begin{aligned}
s_t^n &= \frac{\delta(1+r_t)(\frac{\xi}{\nu}k_t + s_{t-1}^n) - (\delta z_t \bar{c}_t + \frac{\xi}{\nu}k_{t+1})}{1+\delta} \\
&= (1+r_t)\frac{\left[\alpha\delta\frac{\xi}{\nu} - \alpha\delta z_t \frac{\nu(1-\phi)}{\alpha\nu+(1-\alpha)}(\frac{\xi}{\nu}+1) + \left[\delta z_t - \frac{\xi}{\nu}\right]s(\gamma_t, 1, \nu)\right]k_t + \alpha\delta s_{t-1}^n}{\alpha[1+\delta]}.
\end{aligned}$$

Hence, for any $i, j = 1, ..., N$,

$$\frac{s_t^i}{s_t^j} = \frac{\left[\alpha\delta\frac{\xi}{\nu} - \alpha\delta z_t \frac{\nu(1-\phi)}{\alpha\nu+(1-\alpha)}(\frac{\xi}{\nu}+1) + \left[\delta z_t - \frac{\xi}{\nu}\right]s(\gamma_t, 1, \nu)\right]k_t + \alpha\delta s_{t-1}^i}{\left[\alpha\delta\frac{\xi}{\nu} - \alpha\delta z_t \frac{\nu(1-\phi)}{\alpha\nu+(1-\alpha)}(\frac{\xi}{\nu}+1) + \left[\delta z_t - \frac{\xi}{\nu}\right]s(\gamma_t, 1, \nu)\right]k_t + \alpha\delta s_{t-1}^j}.$$

Since, by (28), $\alpha\delta\frac{\xi}{\nu} - \alpha\delta z_t \frac{\nu(1-\phi)}{\alpha\nu+(1-\alpha)}(\frac{\xi}{\nu}+1) + \left[\delta z_t - \frac{\xi}{\nu}\right]s(\gamma_t, 1, \nu) > 0$, we obtain (29).

2) Now let $\gamma_t > \gamma^*(\nu)$. Suppose that $s_{t-1}^j > s_{t-1}^i > 0$ and $s_t^i > 0$. Taking account of Claim 1 and (27), for $n = i, j$, we have

$$\begin{aligned}
s_t^n &= \frac{\delta(1+r_t)(\frac{\xi}{\nu}k_t + s_{t-1}^n) - (\delta z_t \bar{c}_t + \frac{\xi}{\nu}k_{t+1})}{1+\delta} \\
&= (1+r_t)\frac{\left[\alpha\delta\frac{\xi}{\nu} - \alpha\delta z_t \frac{\nu(1-\phi)}{\alpha\nu+(1-\alpha)}(\frac{\xi}{\nu}+1) + \left[\delta z_t - \frac{\xi}{\nu}\right]s(\gamma_t, m(t), \nu)\right]k_t + \alpha\delta s_{t-1}^n}{\alpha(1+\delta)}.
\end{aligned}$$

and hence

$$\frac{s_t^i}{s_t^j} = \frac{\left[\alpha\delta\frac{\xi}{\nu} - \alpha\delta z_t\frac{\nu(1-\phi)}{\alpha\nu+(1-\alpha)}(\frac{\xi}{\nu}+1) + \left[\delta z_t - \frac{\xi}{\nu}\right]s(\gamma_t, m(t), \nu)\right]k_t + \alpha\delta s_{t-1}^i}{\left[\alpha\delta\frac{\xi}{\nu} - \alpha\delta z_t\frac{\nu(1-\phi)}{\alpha\nu+(1-\alpha)}(\frac{\xi}{\nu}+1) + \left[\delta z_t - \frac{\xi}{\nu}\right]s(\gamma_t, m(t), \nu)\right]k_t + \alpha\delta s_{t-1}^j}.$$

Since, by (28), $\alpha\delta\frac{\xi}{\nu} - \alpha\delta z_t\frac{\nu(1-\phi)}{\alpha\nu+(1-\alpha)}(\frac{\xi}{\nu}+1) + \left[\delta z_t - \frac{\xi}{\nu}\right]s(\gamma_t, 1, \nu) < 0$, we obtain $s_t^i/s_t^j < s_{t-1}^i/s_{t-1}^j$. □

**Lemma 2** $\gamma_t \gtreqless \gamma^*(\nu) \Rightarrow \gamma_{t+1} \lesseqgtr \gamma_t$

**Proof.** Lemma 2 follows directly from Lemma 1.

**Proof of Theorem 2.** 1) If $\gamma(s_{-1}^1, ..., s_{-1}^N) < \gamma^*(\nu)$, it is sufficient to refer to Claims 1 and 2 and Lemma 2.

2) Suppose that $\gamma(s_{-1}^1, ..., s_{-1}^N) > \gamma^*(\nu)$. Then by Lemma 2 the sequence $(\gamma^t)_{t=0}^\infty$ is non-decreasing and hence converges to some $\widetilde{\gamma} > \gamma^*(\nu)$. Therefore, by Claim 3, the sequence $(M(t))_{t=0}^\infty$ is non-increasing. It follows that there are $M > 0$ and $T$ such that $M(t) = M$ for $t \geq T$. We need to show that $M = L$.

Assume that $M > L$. Taking into account Claim 1, $\lim_{t\to\infty} k_t = k(\widetilde{\gamma}, M/N, \nu)$.

Let $j \leq L$ and $L < i \leq M$. From Lemma 1, the sequence $(s_t^j/k_{t+1})_{t=0}^\infty$ is increasing and the sequence $(s_t^i/s_t^j)_{t=0}^\infty$ is decreasing. Therefore, the sequence $(s_t^j)_{t=0}^\infty$ converges to some $\widetilde{s}^j > 0$ and the sequence $(s_t^i)_{t=0}^\infty$ converges to some $\widetilde{s}^i$. Moreover, $\widetilde{s}^i < \widetilde{s}^j$, which is impossible. Indeed, by Claim 1, (13) and (27), for $n = 1, ..., M$,

$$\begin{aligned} s_t^n &= \frac{\delta(1+r_t)(\frac{\xi}{\nu}k_t + s_{t-1}^n) - (\delta z_t\bar{c}_t + \frac{\xi}{\nu}k_{t+1})}{1+\delta} \\ &= (1+r_t)\frac{\left[\alpha\delta\frac{\xi}{\nu} - \alpha\delta z_t\frac{\nu(1-\phi)}{\alpha\nu+(1-\alpha)}(\frac{\xi}{\nu}+1) + \left[\delta z_t - \frac{\xi}{\nu}\right]s(\gamma_t, \frac{M}{N}, \nu)\right]k_t + \alpha\delta s_{t-1}^n}{\alpha(1+\delta)}. \end{aligned}$$

hence for both $n = j$ and $n = i$, $\widetilde{s}^n$, we must have

$$\widetilde{s}^n = (1+\widetilde{r})\frac{\left[\alpha\delta\frac{\xi}{\nu} - \alpha\delta\widetilde{z}\frac{\nu(1-\phi)}{\alpha\nu+(1-\alpha)}(\frac{\xi}{\nu}+1) + \left[\delta\widetilde{z} - \frac{\xi}{\nu}\right]s(\widetilde{\gamma}, \frac{M}{N}, \nu)\right]k_t(\widetilde{\gamma}, \frac{M}{N}, \nu) + \alpha\delta\widetilde{s}^n}{\alpha(1+\delta)},$$

where $\widetilde{z} = \frac{\widetilde{\gamma}}{1+\widetilde{\gamma}}$ and $1 + \widetilde{r} = \alpha\left[k(\widetilde{\gamma}, M/N, \nu))\right]^{\alpha-1}$.

**Proof of Propositions 2 and 3.** These propositions follow from Theorem 2.

dam: Amsterdam Institute for Advanced Labour Studies. Available at: http://www.gini-research.org/system/uploads/546/original/90.pdf (accessed on 23/02/2021).

Pham, T.K.C. (2005) Economic growth and status-seeking through personal wealth. *European Journal of Political Economy*, 21, 407 – 427.

Piketty, T., Saez, E. (2013) A theory of optimal inheritance taxation. *Econometrica*, 81 (5), 1851 – 1886.

Piketty, T., Zucman, G. (2015) Wealth and inheritance in the long run. In: Atkinson, A. B., and F. Bourguignon (Eds), *Handbook of Income Distribution,* Elsevier, Vol. 2, 1303 – 1368.

Postlewaite, A. (2011) Social norms and preferences. In: Benhabib, J., A. Bisin, and M. Jackson (Eds), *Handbook for Social Economics*, Amsterdam: North-Holland, Vol. 1A, 31 – 67.

Saez, E., Zucman, G. (2019) Progressive wealth taxation. *Brookings Papers on Economic Activity*, BPEA Conference Draft, September 5-6.

Schneider, S. M. (2019) Why income Inequality is dissatisfying – Perceptions of social status and the inequality-satisfaction link in Europe. *European Sociological Review*, 35, 409 – 430.

Truyts, T. (2010) Social status in economic theory. *Journal of Economic Surveys*, 24, 137 – 169.

Veblen, T. (1899) *The theory of the leisure class*. New York, NY: Macmillan.
41

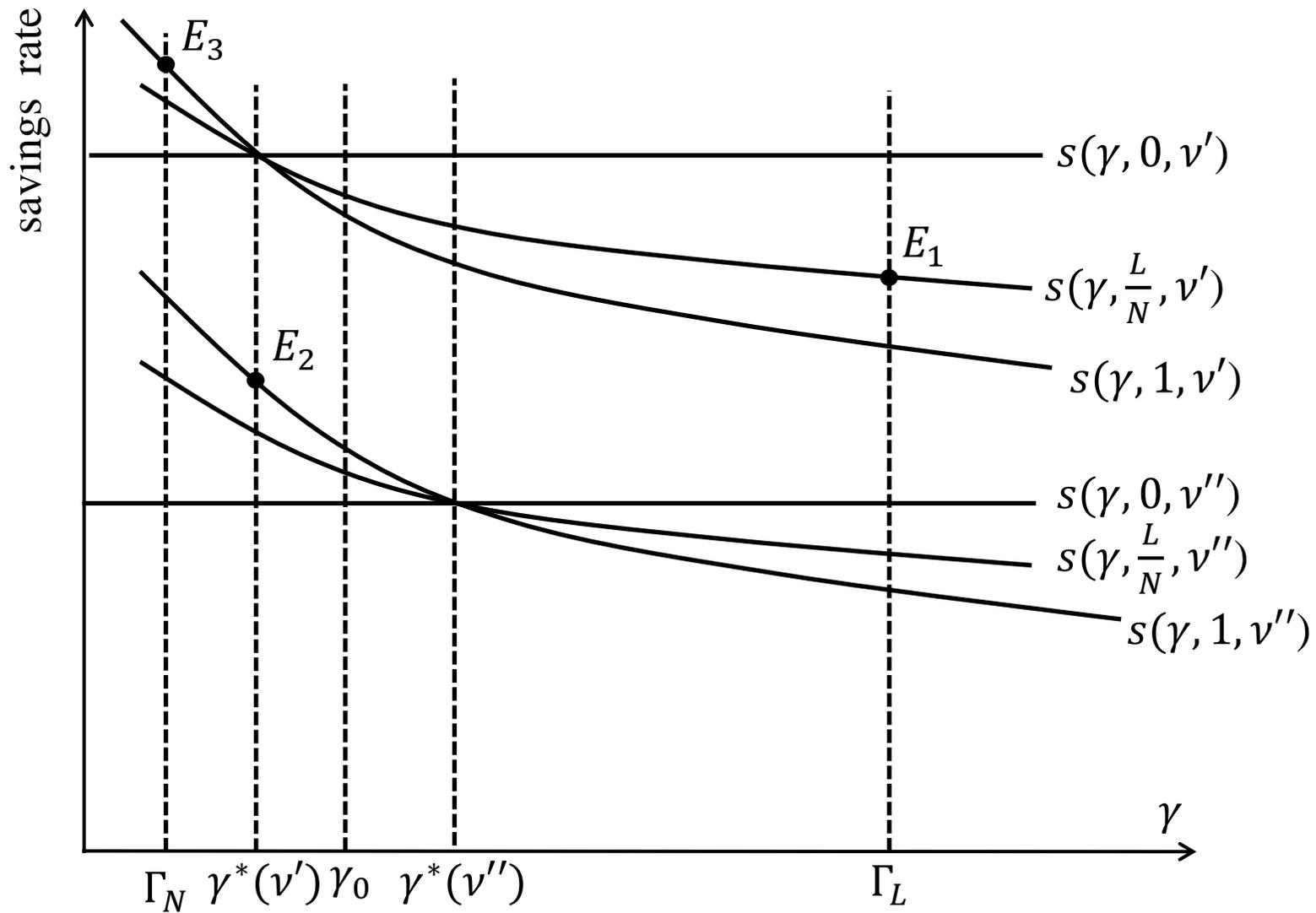

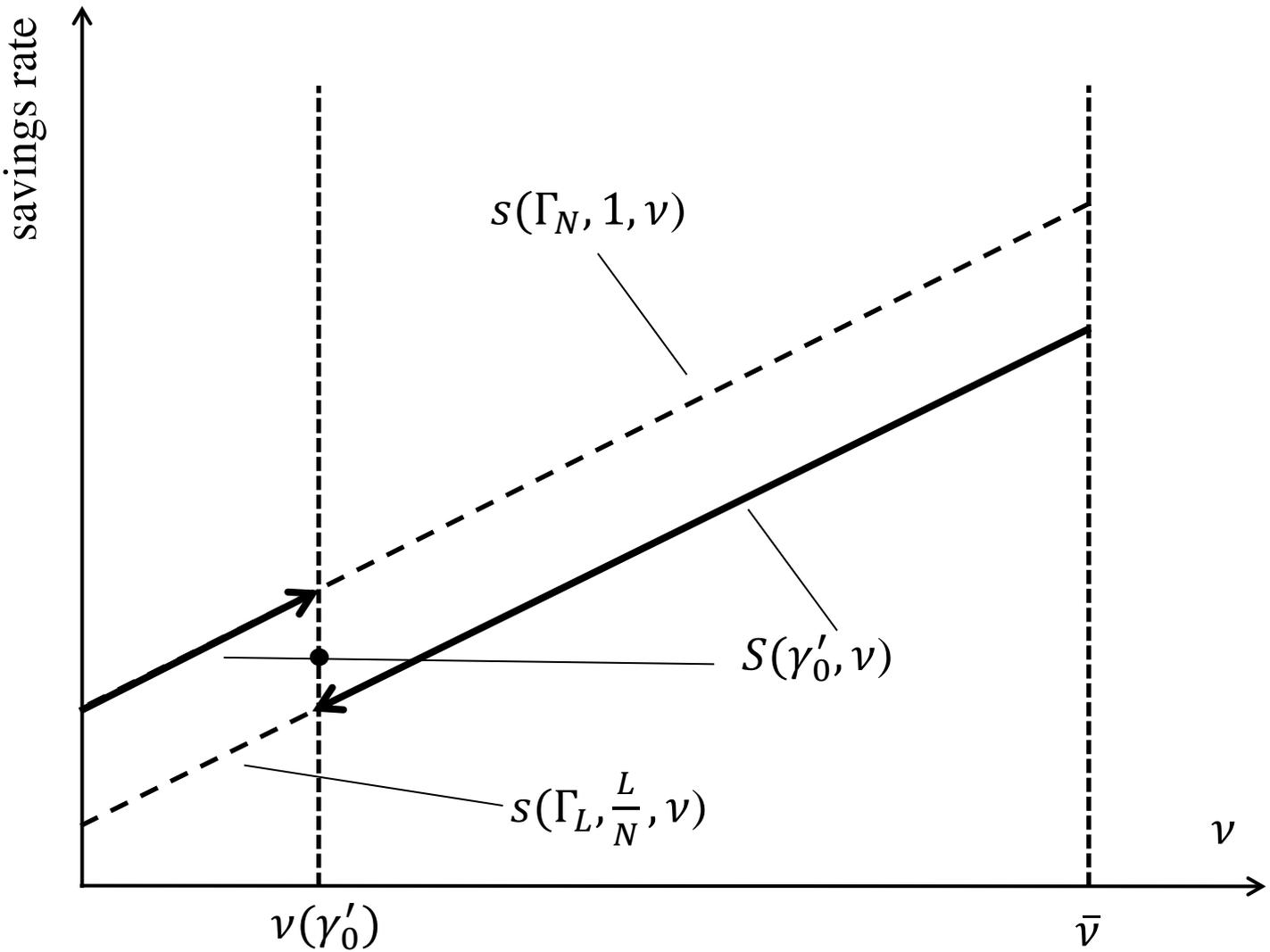
(a)

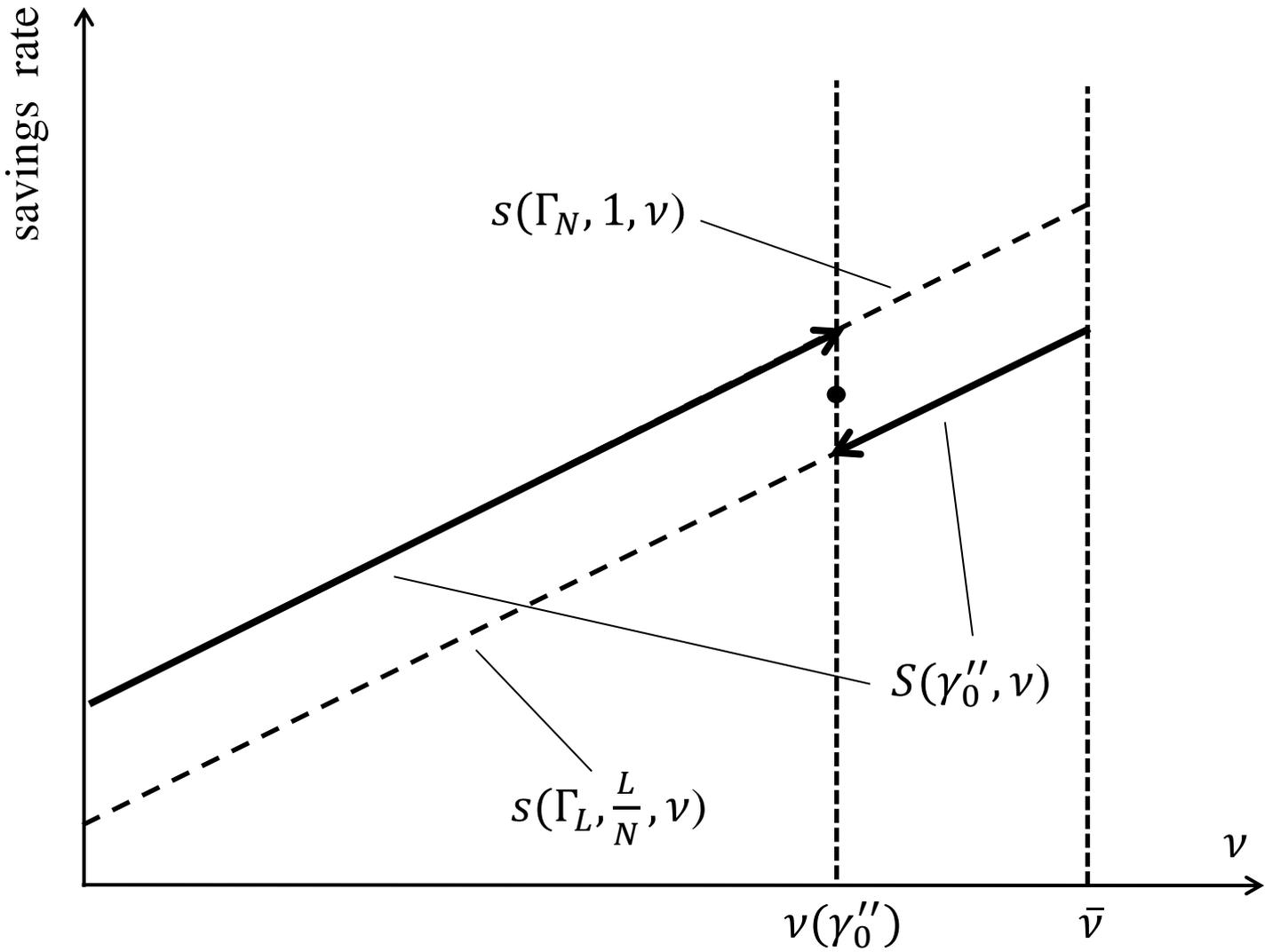

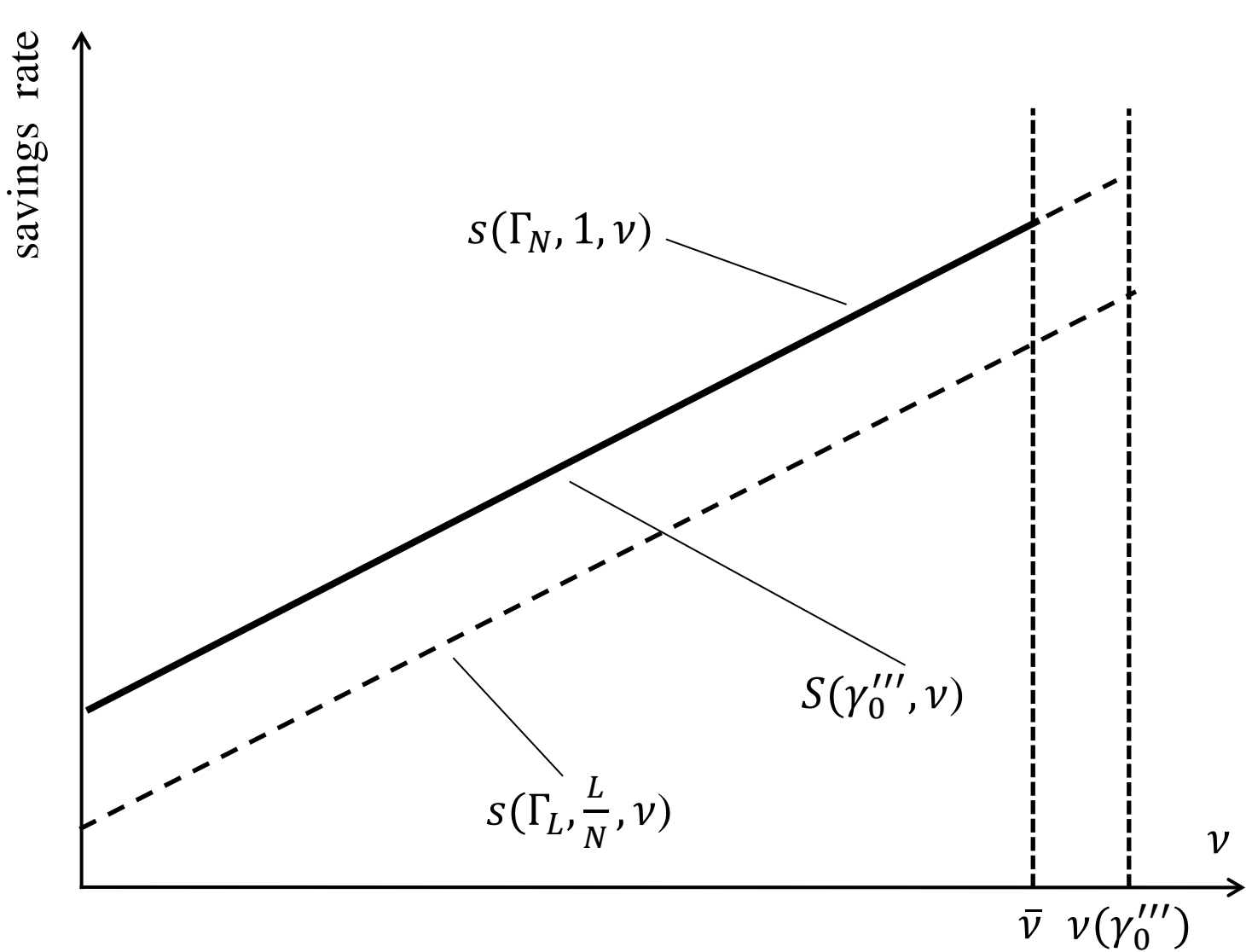

(c)